\documentclass[aps,pra,floatfix,showpacs,noshowkeys,epsfig,graphics,natbib]{revtex4-1}
\usepackage{graphicx}
\usepackage{amsmath}
\usepackage{amsfonts}
\usepackage{amssymb}

\newcommand\fverb{\setbox\fverbbox=\hbox\bgroup\verb}
\newcommand\fverbdo{\egroup\medskip\noindent%
            \fbox{\unhbox\fverbbox}\ }
\newcommand\fverbit{\egroup\item[\fbox{\unhbox\fverbbox}]}
\newbox\fverbbox

\newcommand{\newc}{\newcommand}
\newc{\beq}    {\begin{equation}}
\newc{\eeq}    {\end{equation}}

\newc{\beqa}    {\begin{eqnarray}}
\newc{\eeqa}    {\end{eqnarray}}
\newc{\bs}    {\section}
\newc{\no}    {\\ \nonumber}

\newc{\st}    {\stackrel}

\begin{document}

\title{  Zero Cosmological Constant
and Nonzero   Dark Energy from Holographic Principle }
\author{Jae-Weon Lee}\email{scikid@jwu.ac.kr}
\affiliation{ Department of energy resources development,
Jungwon
 University,  5 dongburi, Goesan-eup, Goesan-gun Chungbuk Korea
367-805}

\date{\today}

\begin{abstract}
It is shown that the first law of thermodynamics
and the holographic principle applied to an arbitrary large
cosmic causal horizon
naturally demand the zero cosmological constant
and non-zero  dynamical dark energy
in the form of the holographic dark energy.
Semiclassical analysis shows that the holographic dark energy has
 a parameter $d=1$ and an equation of state
comparable to  current observational data, if the entropy of the horizon
saturates the Bekenstein-Hawking bound. This result
indicates that quantum field theory should be modified at large scale
to explain dark energy.
The relations among dark energy, quantum  vacuum energy and entropic gravity are
also discussed.
\end{abstract}

\maketitle

Type Ia supernova (SN Ia) observations~~\cite{riess-1998-116,perlmutter-1999-517},
the Sloan Digital Sky Survey (SDSS)
~\cite{gong:043510,zhang:043524,1475-7516-2004-08-006,seljak-2006-0610}
and cosmic microwave background observations~\cite{Komatsu:2010fb}
all indicate
that
the  current universe is expanding at an accelerating rate.
The expansion can be explained if there is a
negative pressure fluid called
 dark energy of which pressure $p_{DE}$ and  energy density $\rho_{DE}$ satisfy
$w_{DE}\equiv p_{DE}/\rho_{DE}<-1/3$.
Being one of the most important unsolved
puzzles  in modern physics and cosmology, the dark energy problem consists of three sub-problems ~\cite{CC};
why it is so small, nonzero, and comparable to the critical density at the present.

We also need to  explain why the cosmological constant $\Lambda$ is so small
 or exactly zero.
Solving this problem is not an easy task, because quantum field theory (QFT) predicts  huge zero point energy
that  can
play a role of $\Lambda$.
It is very hard to reconcile the great success of
QFT at small scales with
this failure of QFT to explain dark energy.
There are already many works on this problem~\cite{Bousso:2007gp,Guberina:2005fb,Erdem:2004yd}, however,
the problem seems to be far from a solution.

In this paper, it is suggested that if the holographic principle holds for a cosmic causal horizon,
the cosmological constant should be exactly zero and
 there should be holographic dark energy (HDE) consistent with  the recent observational data.
The holographic principle~\cite{hooft-1993}
 is a  conjecture  claiming
 that all of  the information in a region  can be described by the physics
  at the boundary
 of the region  and that the maximal number of degrees of freedom in the region
  is proportional to its surface area
 rather than the volume.
 More specifically, it was conjectured that the Bekenstein-Hawking Entropy
\beq
S_{BH}=\frac{ c^3~A}{4G\hbar}
\eeq
 is the information bound that a region of space with
 a surface area $A$ can contain~\cite{Bekenstein:1993dz}.

Based on black hole physics Cohen et al~\cite{PhysRevLett.82.4971} proposed   that
the total energy in a region  can not be larger than
that of a black hole of that size. Therefore, if the region has a size $r_h=O(H^{-1})$,
the vacuum energy density is bounded as
$\rho_\Lambda\leq O(M_P^2 H^2)$, where
$H=da/adt$ is the Hubble parameter with the scale factor $a$, and $M_P=\sqrt{\hbar c/8 \pi G}$
is the reduced Planck mass.
  Interestingly, saturating the bound gives HDE
comparable to the observed  dark energy density $\rho_\Lambda\sim
10^{-10} eV^4$.
However, Hsu~\cite{hsu} pointed out that with the Hubble horizon HDE behaves like
matter rather than dark energy. Li~\cite{li-2004-603} suggested that
 using  the future event horizon as IR cutoff we can solve this problem.

Recently, based on the holographic principle
 Verlinde ~\cite{Verlinde:2010hp} and Padmanabahan ~\cite{Padmanabhan:2009kr}
proposed a remarkable idea linking gravity to entropy, which brings out
  many follow-up studies
~\cite{Zhao:2010qw,Cai:2010sz,Myung:2010jv,Liu:2010na,Tian:2010uy,Diego:2010ju,Pesci:2010un,Vancea:2010vf,Konoplya:2010ak,Culetu:2010ua,Zhao:2010vt,Ghosh:2010hz}.
Verlinde derived the Newton's equation and the Einstein equation by assuming
that energy inside a holographic screen is
the  equipartition energy $E_h\sim T_h N$ for the screen
with the temperature $T_h$ and the number of bits $N$.

On the other hand, in a series of works ~\cite{myDE,Kim:2007vx,Kim:2008re,Lee:2010bg}, Lee et al.  suggested  that the energy of gravitational systems
could be explained by  considering information loss at causal horizons.
For example, we pointed out that
a cosmic
causal horizon with a radius $r\sim O(H^{-1})$ has
  temperature $T_h\sim 1/r$,   entropy
$S_h\sim r^2$
and
a kind of  thermal
energy $E_{h}\sim T_h S_h\sim r$,
which can be dark energy ~\cite{Li:2011sd}.
 This     dark energy, dubbed `quantum informational dark energy'~\cite{Lee:2007ky}
or `entanglement dark energy'~\cite{myDE} by the authors, is  similar to the entropic dark energy
  based on
the Verlinde's idea~\cite{Li:2010cj,Zhang:2010hi,Wei:2010ww,Easson:2010av}.
It was also suggested that black hole mass and the Einstein equation itself
 can be derived from the relation
 $dE_h=k_B T_h dS_h$, that  might have a quantum information theoretic origin~\cite{Kim:2007vx}.
Similarities between this theory and Verlinde's theory were investigated in  ~\cite{Lee:2010bg,Lee:2010fg}.

In this paper we assume that the holographic principle and
the following first-law like definition of the horizon energy
\beq
\label{eenergy}
dE_{h}\equiv k_B T_h dS_{h},
\eeq
hold for a  cosmic causal horizon such as the cosmic event horizon or the apparent horiozn.
This energy could be the equipartition energy~\cite{Verlinde:2010hp},
energy from  Landauer's principle
associated with information loss at the horizon~\cite{myDE}
or simply the energy defined by the Clausius relation.
Inspired by the  entropic~\cite{Verlinde:2010hp, Padmanabhan:2009vy} or quantum information theoretic~\cite{myDE, Lee:2010bg}
interpretation of gravity
we take the holographic principle and the horizon energy in Eq. (\ref{eenergy})
as guiding principles for dark energy study.

Let us first recall the cosmological constant problem in the context of QFT.
The (classical) time independent cosmological constant $\Lambda_c$  appears in the gravity action as
\beq
\label{a}
S=\int d^4x \sqrt{-g}(R-2\Lambda_c).
\eeq
Since the energy-momentum tensor  $T_{\mu\nu}$
for the vacuum fluctuation  $\langle T_{\mu\nu} \rangle$  is usually proportional to a spacetime metric (See
for example ~\cite{RevModPhys.57.1}),
$\langle T_{\mu\nu} \rangle$ has been regarded as
 a candidate for the cosmological constant and
 dark energy. To calculate
its expectation value
one usually integrates the   zero point energy $\hbar \omega/2$ for each mode of  quantum fields
in a flat spacetime.
Thus, the energy density of the quantum vacuum is approximately given by
\beq
\label{rhovac}
\rho_{q}=\langle T_{00} \rangle \sim
\int^{k_U}_{k_I} \hbar  \omega   d^3 k \sim k_U^4,
\eeq
where $k_U\sim M_P$ is a UV-cutoff and  $k_I\sim 1/r$ is an  IR-cutoff.
Unfortunately, as is well known, for $k_U\sim M_P$, the estimation gives
 $\rho_{q}\sim M^4_P \sim 10^{109} eV^4$ which
is too large to explain the
observed dark energy density $\rho_{DE}\sim 10^{-12} eV^4$.
On the other hand, if we subtract this zero point energy
   to calculate the renormalized vacuum energy for the universe, we  obtain $\rho_q\sim H^4$,
which is too small compared to the observed dark energy.

It is  often argued that after taking the vacuum expectation of  quantum fields,
the Friedmann equation
\beq
H^2=\frac{8\pi G \rho_m}{3}-\frac{k_c c^2}{R}+\frac{\Lambda c^2}{3},
\eeq
gets an additional
 constant contribution $\Lambda_q=\rho_q/M_P^2 c^2=\langle T_{00}\rangle/M_P^2 c^2$ from the vacuum quantum fluctuation $\rho_{q}$ in
Eq. (\ref{rhovac}).
(Here,  $k_c$ is the spatial curvature parameter,
which we will set zero for simplicity, and
$\rho_m$ is the matter energy density.)
Thus, the total cosmological constant is
$
\Lambda=\Lambda_c +\Lambda_q,
$
and
the total vacuum energy density is given by
\beq
\label{rhovac2}
\rho_{vac}= M_P^2 c^2 (\Lambda_c  + \Lambda_q  ).
\eeq
Without a fine tuning it seems to be almost impossible for two terms to cancel
each other to result in the tiny observed upper bound for  the cosmological constant.
This is the essence of the cosmological constant problem.

Then, in the context of QFT, from where could horizon energy $\rho_h$  arise?
Recall that $\rho_{q}$ in Eq. (\ref{rhovac}) was estimated in a flat spacetime.
However, for a curved spacetime, after a
Bogoliubov  transformation there appear excited states in addition to the vacuum.
 The normal ordered
 quantum vacuum energy (i.e., with the subtraction of the zero point energy) in a curved spacetime with the UV and the IR cutoffs
often has a term   in the form of $\rho_h$.
One can do a volume integral of $\hbar \omega$ with
  the Bogoliubov  coefficient $\beta_k$ for the quantum fields in a curved spacetime
   to obtain $\rho_h$.
  For example,  using the result in ~\cite{UV}, it was shown in ~\cite{Lee:2008vn} that  $\rho_{vac}$   for
   the de Sitter universe  contains an extra term
\beq
\rho'_{vac}\sim \int^{k_U}_{k_I} d^3k~ \hbar \omega |\beta_k|^2 \sim k_U^2  H^2
\eeq
in addition to the usual zero point energy, where
$\beta_k\sim H/k$ .
If we choose
 $M_P$ as the UV-cutoff $k_U$, this extra term gives  $\rho'_{vac}\sim M_P^2  H^2 \sim \rho_h$.
Thus, it is possible that $\rho_h$  is actually
the average quantum fluctuation energy above the zero point vacuum  energy of the curved spacetime in the bulk~\cite{Padmanabhan:2008if}.
 This dark energy may be also identified to be the energy of cosmic Hawking radiation ~\cite{Lee:2008vn}.
However, this calculation still can not explain why we can ignore
the zero point energy in the bulk. It seems that there is no
plausible way to overcome this  difficulty, as long as we rely on
the conventional QFT. We need another fundamental ingredient to
solve this problem.

\begin{figure}[htbp]
\includegraphics[width=0.4\textwidth]{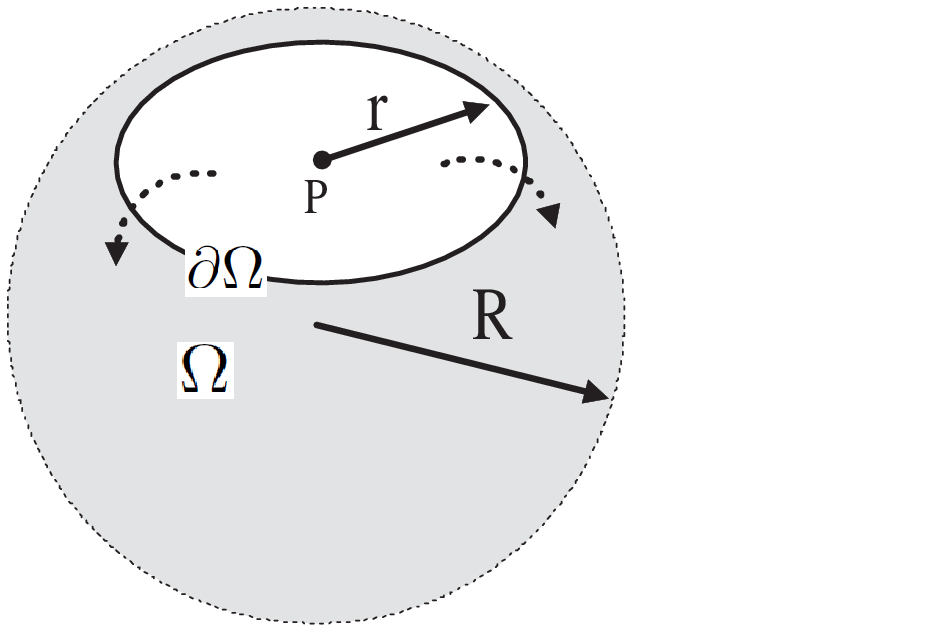}
\caption{
A cosmological causal horizon $\Sigma$ with a radius $r$, temperature $T_h$, and entropy $S_h$
  has a dark energy $E$ given by $dE=k_B T_hdS_h\sim r$.
  The holographic principle for arbitrary $r$ demands the cosmological constant to be exactly zero.
   }
\end{figure}

Alternatively, we can take not the bulk QFT but
the holographic  principle as a postulate and  describe the bulk physics
using only the DOF on the horizon.
In this holographic context, to estimate the  bulk energy density we can treat the quantum fields on the horizon
as a collection of oscillators on the spherical surface with  a lattice constant of order $O(k_U^{-1})$.
Then, to obtain $\rho_{h}$ we have to sum the zero-point energy of the oscillators
with frequency $\omega$ on the horizon surface
rather than those in the bulk. This rough estimation results in the HDE density, because
\beq
\label{rhoholo}
\rho_{h}\sim \Sigma_i \frac{\hbar \omega}{volume} \sim \Sigma_i \frac{\hbar \omega}{r^3} \sim  \left(\frac{r}{k_U^{-1}}\right)^2\frac{\hbar \omega}{r^3}
\sim  \frac{M_P^2}{r^2}\sim M_P^2 H^2,
\eeq
where  $\Sigma_i$ represents a summation over the horizon oscillators  with the temperature $T_h$,
and the number of oscillators are proportional to the horizon area $\sim ({r}/{k_U^{-1}})^2$.
At the last step we used the equipartition approximation $\hbar \omega \sim k_B T_h\sim 1/r$.
Note that this is just an order of magnitude estimation for comparison, and more accurate
solution requires a careful calculation with an appropriate horizon.

 This result indicates that the bulk QFT overestimates the independent DOF in the bulk and
 the true vacuum energy of the bulk could be
 the zero point energy of the boundary DOF on the horizon, which is of order of  the normal-ordered bulk vacuum energy in the conventional QFT.
What gives the small  HDE could be
 the smallness of the number of independent DOF in the bulk.
This redundancy of the bulk DOF can
 explain  why we cannot obtain the correct dark energy density by simply calculating
 the zero point energy  of the bulk. In short, QFT is not a complete theory at the cosmic scale.

%

We need to calculate the horizon energy $E_h$ as the vacuum energy of the universe
without using QFT.
 Let us consider a causal cosmic horizon with a radius $r$,
 having generic holographic entropy
\beq
\label{Sh}
S_{h}= \frac{\eta  c^3 r^2}{G \hbar},
\eeq
and temperature
\beq
\label{Th}
T_h=\frac{\epsilon \hbar c}{k_B r},
\eeq
with constants $\eta$ and $\epsilon$ (See Fig. 1).
(Note that these quantities contatin $\hbar$ and
are usually derived by semiclassical calculations.)
In this case the universe is similar to a big black hole with an expanding horizon.
For  the Bekenstein entropy $\eta=\pi$, and  the Hawking-Gibbons temperature $\epsilon=1/2\pi$.
By assuming the first law and integrating $dE_{h}$ on the isothermal surface $\Sigma$
of the causal horizon with Eqs. (\ref{Sh}) and (\ref{Th}), we obtain
the horizon energy
 \beq
 \label{Eh}
E_{h}=\int_{\Sigma}  dE_{h}= k_B T_{h} \int_{\Sigma}  dS_{h}=\frac{\eta \epsilon c^4 r  }{G}.
\eeq
Then, the
energy density due to $E_h$  is given by
 \beq
\label{rhoh}
\rho_{h}=\frac{3 E_{h}}{4 \pi r^3}
=\frac{6 \eta \epsilon c^3    M_P^2}{\hbar   r^2} \equiv \frac{3 d^2 c^3 M_P^2  }{ \hbar  r^2 },
\eeq
which  has  the form  of the
holographic dark energy~\cite{li-2004-603}.
Note that this semiclassical derivation of HDE is different from the usual derivation based on UV-IR relations.
$\rho_h$  here  corresponds to the estimation of the surface vacuum energy in Eq. (\ref{rhoholo}).
This kind of dark energy was also derived in terms of entanglement energy~\cite{myDE}
and quantum entanglement force~\cite{Lee:2010fg}.
 From the above
equation  we  immediately obtain a  formula for the constant
 \beq
\label{d1}
 d=\sqrt{{2\eta\epsilon}},
 \eeq
which is the important parameter determining the nature of HDE.
If $S_{h}$ saturates the Bekenstein bound
and $T_h$ is the Hawking-Gibbons temperature
${ \hbar c}/{2\pi k_B r}$, then $\eta\epsilon=1/2$ and $d=1$.
Thus,
the holographic principle applied to a cosmic causal horizon naturally leads to HDE
with $d=1$  ~\cite{Lee:2010fg}, which  is favored by observations and theories~\cite{Huang:2004ai,zhang-2007}.
There are few works on fixing $d$ value.
Li found $d=1$ by assuming the universe as a $classical$ black hole~\cite{li-2004-603}.
On the contrary in this paper $d$ value is obtained by considering the semiclassical quantities.


Let us turn to the cosmological constant problem in this context.
From the holographic viewpoint, it is very simple to see why the cosmological constant $\Lambda$ should be zero.
If we apply the holographic principle  and the definition of the horizon energy (Eq. (\ref{Eh})) to the cosmic horizon,
the bulk vacuum energy  density
  $\rho_{vac}$ in Eq. (\ref{rhovac2})  should be smaller than the horizon energy density $\rho_h$ in Eq. (\ref{rhoh}).
  Since the principle is one of our starting postulates, the principle
  should hold strictly even for arbitrary large $r$, and hence, the  vacuum energy $E_{\Lambda}$
   proportional to $\Lambda  r^3$ is problematic. It clearly violates the holographic principle for large $r$,
   where vacuum energy is dominant.
    According to the principle and the first law of thermodynamics with $T_h\propto 1/r$,
   the total horizon energy $E_h$ is proportional to $r$.
   For $r > r_c\equiv \sqrt{\frac{3d^2}{\Lambda}}$, $E_\Lambda>E_h$ and the holographic principle can be violated.
  Thus, the principle holds true for $arbitrary$ $r$ only if the   cosmological constant $\Lambda$ is exactly zero.
  Note that this argument holds for arbitrary small coefficient of $r^3$ term  in $E_\Lambda$ as long as $r$ can increase
  infinetely. As the universe expands,  the inequality
   $E_\Lambda\le E_h$ would be violated eventually.
   For example, if the cosmological constant is the dark energy, constant $\rho_\Lambda$ is  about the present  critical density
   $\rho_c=3 H^2 M_P^2/8\pi\propto 1/t^2$ and within a few Hubble times  $\rho_\Lambda r^3$ will
   exceed $\rho_h r^3$.
  Thus, we can say that the holographic principle
   insists that the  cosmological constant is zero (i.e., $\omega_{DE}\neq -1$).
   (Here, we have excluded an implausible case that $\Lambda_c$ and a constant part of the
  quantum contribution miraculously  cancel each other to result in $\rho_{vac}\sim 1/r^2$.)

This solution to the cosmological constant problem has its own cost.
 At the large cosmic scale,  we have to abandon QFT  and accept the holographic principle and the dark energy problem
becomes much easier.
As long as the  principle holds, the argument about the zero cosmological constant
would be valid. Since the principle also solves the other subproblems about dark energy,
this approach seems to be promising.

Furthermore, the solution above  is free from  some
difficulties often encountered by other approaches such as
infrared or ultraviolet modifications of gravity,
adjusting initial conditions, or dynamical attractor mechanisms (See \cite{ccreview} for example.).
They failed to explain  both of the early small universe
and current large universe and why the QFT vacuum loops or
cosmological phase transitions did not curl up the universe.
Let us discuss these facts in detail.

First, our approach based on the holographic principle suggests that
the energy density from the vacuum loop energy for a quantum field with a UV cutoff
energy scale $M$ is  $\rho_M=O(M^2 H^2)\ll O(M_P^2 H^2)$   not of $O(M^4)$. Since the Friedmann equation
is $\rho_{tot}= 3H^2 M_P^2$, the vacuum loop energy is not a dominant contribution to $\rho_{tot}$
unless $M\simeq M_P$.
Second, our approach could also avoid the issue related to the cosmological phase transitions.
For example, consider a phase transition of the scalar field field $\phi$ with a thermal effective  potential $V(\phi,T)$,
showing the transition at the temperature $T=T_c$.
Then, in conventional approaches even if $\rho_\Lambda$ was set to be $0$ before the transition, during the phase transition
the potential  could generate temporally the energy difference between the false vacuum and the true vacuum,
which is of $-O(T_c^4)=-O(M^2_P H^2)$, where $H$ is the Hubble parameter at the transition.
If the absolute value of energy matters,
this energy difference  could act as a negative cosmological constant and  make the universe  rapidly collapse.
However, in our theory there is always positive  $O(M^2_P H^2)$ dark energy that could
cancel the negative energy term and prevent the collapse.
Third, unlike the dynamical attractor theories, our theory does not directly rely on the contributions of matters
 to energy-momentum tensor and hence we do not  need a feedback mechanism adjusting $\rho_\Lambda$ to
precision $10^{-120}$ as long as the horizon radius is $O(H^{-1})$.

One can easily see the similarity between our theory and entropic gravity.
In entropic gravity the horizon energy is given by the equipartition law
 $E_h= N T_h/2$,  which is essentially equivalent to
 our dark energy $E_h= \int T_h dS$, because $S\sim N$ in general.
Following \cite{Lee:2010fg} and \cite{Easson:2010av} one can also obtain an entropic force
for the dark energy
\beq
\label{entforce}
F_{h}\equiv \frac{dE_{h}}{dr}=\frac{c^4 \eta\epsilon}{ G},
\eeq
which could be also identified as  a `quantum entanglement force' as in \cite{Lee:2010fg},
if $S_h$ is the entanglement entropy.

Let us compare predictions of our theory with observational data.
From $\rho_{DE}=\rho_h$ and a cosmological energy-momentum conservation equation,
one can obtain
an {\it effective}  dark energy pressure ~\cite{li-2004-603} in the bulk
\beq
\label{p}
p_{DE}=\frac{d(a^3\rho_h(r))}{-3 a^2 da },
\eeq
from which one can derive the equation of state.
\begin{figure}[htbp]
\includegraphics[width=0.4\textwidth]{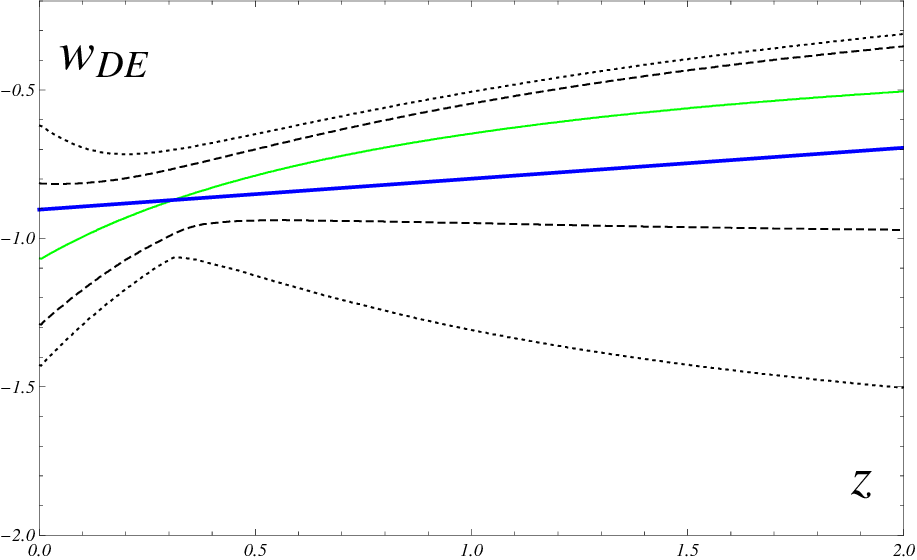}
\caption{ (Color online) Theoretical evolution of the dark energy equation of state (the blue thick line)
 $w_{DE}$ versus the redshift $z$
 compared to the   observational
constraints
 (Data extracted from Fig. 2 in ~\cite{Gong:2006gs}).
 The
green thin line represents the best fit.
 The dashed lines and the dotted lines
 shows  $1\sigma$ and $2\sigma$ errors, respectively.
 \label{Fig2} }
\end{figure}

At this point
 we need to choose a horizon among various cosmological horizons
such as an apparent horizon, a Hubble horizon, and a future event horizon.
In the simplest case, only the event horizon can result in the accelerating universe~\cite{li-2004-603}.
Thus, from now on we assume the  case that the causal horizon is the cosmic event horizon. For this case
one can find  the equation of state for holographic dark energy as a function of the redshift $z$ as shown in Ref. \cite{li-2004-603}.
Fig. 1 compares this prediction with $d=1$ to the observational data obtained from the 182 gold SN Ia data,
the baryon acoustic oscillation, SDSS, and the 3-year Wilkinson Microwave
Anisotropy Probe (WMAP) data.
One can also find that
the equation of state  ~\cite{li-2004-603,1475-7516-2004-08-006}
$
\label{omega0}
w_0 =-\frac{1}{3} \left(1+\frac{2\sqrt{\Omega^0_\Lambda} }{d}\right),
$
and
its  change rate at  the present $w_1$
with
$w_{DE} (a) \simeq w_0+w_1 (1-a)$.
Here the current dark energy density parameter $\Omega^0_\Lambda\simeq 0.73$.

 For $d=1$ these equations give $w_0=-0.903$ and $w_1=0.104$.
According to  WMAP 5-year data ~\cite{Komatsu:2008hk},
$w_0 = -1.04\pm 0.13$
  and  $w_1 = 0.11\pm 0.7 $.
  WMAP 7-year data with
 the  baryon acoustic oscillation, SN Ia, and the Hubble constant
  yields $w_0 = -0.93\pm 0.13$
  and  $w_1 = -0.41^{+0.72}_{-0.71}$ ~\cite{Komatsu:2010fb}.

  If we use an entanglement entropy calculated in \cite{Lee:2010fg} for $S_h$,
one can obtain $d$ slightly different from $1$.
It is also straightforward to study the cases with other horizons such as apparent horizons or Hubble horizons.
We saw that the predictions of our theory well agree with the recent observational data.
Note that although the cosmological constant is most favored by the cosmological observations,
the observational data  still allow dynamical dark energy models.

It was also shown that holographic dark energy models  with an inflation with
a number of e-folds $N_e\simeq 65$
can solve the cosmic coincidence problem~\cite{mycoin,li-2004-603}
thanks to a rapid expansion of the event horizon during the inflation.

I summarize how the holographic principle and the horizon energy
 can solve the dark energy problem.
In this theory
the dark energy density is small due to the holographic principle,
 comparable to the critical density due to the  $O(1/H)$ horizon size,
and non-zero due to  the quantum fluctuation.
The vacuum fluctuation energy is not huge but comparable to the observed dark energy,
 because conventional QFT overestimates the actual independent DOF.
 The holographic principle and the first law of thermodynamics also
 demand that  the cosmological constant is zero, because the nonzero time independent cosmological constant
 is inconsistent with them.

Compared to previous works on HDE, our work has following new features.
First, albeit simple, the dark energy theory in this paper seems to provide us a logically self-consistent explanations to
the all subproblems of the dark energy including the cosmological constant problem.
Second, the parameter $d$ is obtained using semiclassical parameters such as Hawking temperature
incorporating quantum effects to some extent.
Third, the relations among HDE,  QFT vacuum energy and entropic gravity are studied.

 Note that our solution
 is more than a simple transformation of one problem into another one,
 because the formalism we used here is  based not on the conventional QFT but on the holographic principle
 that could allow a reformulation of gravity and quantum mechanics in terms of thermodynamics as recently suggested
 by some authors~\cite{Verlinde:2010hp,Padmanabhan:2008if, myquantum}.
Therefore, there is interesting possibility that these
thermodynamic approaches  could open a new route to understanding
not only dark energy but also the unification of quantum mechanics
and gravity in the future.

\section*{acknowledgments}
This work was supported in part by Basic Science Research Program through the
National Research Foundation of Korea (NRF) funded by the ministry of Education, Science and Technology
(2010-0024761) and the topical research program (2010-T-1) of Asia Pacific Center for Theoretical
Physics.
%

\begin{thebibliography}{10}

\bibitem{riess-1998-116}
A.~G. Riess {\it et~al.}, Astron. J. {\bf 116},  1009  (1998).

\bibitem{perlmutter-1999-517}
S. Perlmutter {\it et~al.}, Astroph. J. {\bf 517},  565  (1999).

\bibitem{gong:043510}
Y. Gong, B. Wang, and Y.-Z. Zhang, Phys. Rev. D {\bf 72},  043510  (2005).

\bibitem{zhang:043524}
X. Zhang and F.-Q. Wu, Phys. Rev. D {\bf 72},  043524  (2005).

\bibitem{1475-7516-2004-08-006}
Q.-G. Huang and Y. Gong, JCAP {\bf 2004},  006  (2004).

\bibitem{seljak-2006-0610}
U. Seljak, A. Slosar, and P. McDonald, JCAP {\bf 0610},  014  (2006).

\bibitem{Komatsu:2010fb}
E. Komatsu {\it et~al.}, arXiv:1001.4538  (2010).

\bibitem{CC}
P.~J.~E. Peebles and B. Ratra, Rev. Mod. Phys. {\bf 75},  559  (2003).

\bibitem{Bousso:2007gp}
R. Bousso, Gen. Rel. Grav. {\bf 40},  607  (2008).

\bibitem{Guberina:2005fb}
B. Guberina, R. Horvat, and H. Stefancic, JCAP {\bf 0505},  001  (2005).

\bibitem{Erdem:2004yd}
R. Erdem, Phys. Lett. {\bf B621},  11  (2005).

\bibitem{hooft-1993}
G. 't~Hooft, {\em Salam-festschrifft} (World Scientific, Singapore, 1993).

\bibitem{Bekenstein:1993dz}
J.~D. Bekenstein, Phys. Rev. {\bf D49},  1912  (1994).

\bibitem{PhysRevLett.82.4971}
A.~G. Cohen, D.~B. Kaplan, and A.~E. Nelson, Phys. Rev. Lett. {\bf 82},  4971
  (1999).

  \bibitem{hsu}
S.~D.~H. Hsu, Phys. Lett. B {\bf 594},  13  (2004).

\bibitem{li-2004-603}
M. Li, Phys. Lett. B {\bf 603},  1  (2004).


\bibitem{Verlinde:2010hp}
E.~P. Verlinde, arXiv:1001.0785  (2010).

\bibitem{Padmanabhan:2009kr}
T. Padmanabhan, arXiv:0912.3165  (2009).

\bibitem{Zhao:2010qw}
L. Zhao, arXiv:1002.0488  (2010).

\bibitem{Cai:2010sz}
R.-G. Cai, L.-M. Cao, and N. Ohta, arXiv:1002.1136  (2010).

\bibitem{Myung:2010jv}
Y.~S. Myung, arXiv:1002.0871  (2010).

\bibitem{Liu:2010na}
Y.-X. Liu, Y.-Q. Wang, and S.-W. Wei, arXiv:1002.1062  (2010).

\bibitem{Tian:2010uy}
Y. Tian and X. Wu, arXiv:1002.1275  (2010).

\bibitem{Diego:2010ju}
M. Diego, arXiv:1002.1941  (2010).

\bibitem{Pesci:2010un}
A. Pesci, arXiv:1002.1257  (2010).

\bibitem{Vancea:2010vf}
I.~V. Vancea and M.~A. Santos, arXiv:1002.2454  (2010).

\bibitem{Konoplya:2010ak}
R.~A. Konoplya, arXiv:1002.2818  (2010).

\bibitem{Culetu:2010ua}
H. Culetu, arXiv:1002.3876  (2010).

\bibitem{Zhao:2010vt}
Y. Zhao, arXviv:1002.4039  (2010).

\bibitem{Ghosh:2010hz}
S. Ghosh, arXiv:1003.0285  (2010).

\bibitem{myDE}
J.-W. Lee, J. Lee, and H.-C. Kim, JCAP08(2007)005;hep-th/0701199  (2007).

\bibitem{Kim:2007vx}
H.-C. Kim, J.-W. Lee, and J. Lee,
Mod.\ Phys.\ Lett.\ A {\bf 25}, 1581 (2010)  [arXiv:0709.3573 [hep-th]].


\bibitem{Kim:2008re}
H.-C. Kim, J.-W. Lee, and J. Lee, JCAP {\bf 0808},  035  (2008).

\bibitem{Lee:2010bg}
J.-W. Lee, H.-C. Kim, and J. Lee, arXiv:1001.5445  (2010).

\bibitem{Li:2011sd}
  M.~Li, X.~D.~Li, S.~Wang and Y.~Wang,
  Commun.\ Theor.\ Phys.\  {\bf 56}, 525 (2011)
  [arXiv:1103.5870 [astro-ph.CO]].

\bibitem{Lee:2007ky}
J.-W. Lee, J. Lee, and H.-C. Kim, {\it Proceedings of the National Institute for Mathematical Science} 8, 1 (2007)
arXiv:0709.0047  (2007).

\bibitem{Li:2010cj}
M. Li and Y. Wang, arXiv:1001.4466  (2010).

\bibitem{Zhang:2010hi}
Y. Zhang, Y.-g. Gong, and Z.-H. Zhu, arXiv:1001.4677  (2010).

\bibitem{Wei:2010ww}
S.-W. Wei, Y.-X. Liu, and Y.-Q. Wang, arXiv:1001.5238  (2010).

\bibitem{Easson:2010av}
D.~A. Easson, P.~H. Frampton, and G.~F. Smoot, arXiv:1002.4278  (2010).

\bibitem{Lee:2010fg}
J.-W. Lee, H.-C. Kim, and J. Lee, arXiv:1002.4568  (2010).

\bibitem{Padmanabhan:2009vy}
T. Padmanabhan, arXiv:0911.5004  (2009).

\bibitem{RevModPhys.57.1}
R.~H. Brandenberger, Rev. Mod. Phys. {\bf 57},  1  (1985).


\bibitem{Huang:2004ai}
Q.-G. Huang and M. Li, JCAP {\bf 0408},  013  (2004).

\bibitem{zhang-2007}
X. Zhang and F.-Q. Wu, arXiv:astro-ph/0701405  (2007).

\bibitem{Lee:2008vn}
J.-W. Lee, H.-C. Kim, and J. Lee, Mod. Phys. Lett. {\bf A25},  257  (2010).

\bibitem{UV}
E. Keski-Vakkuri and M.~S. Sloth, JCAP {\bf 2003},  001  (2003).

\bibitem{Padmanabhan:2008if}
T. Padmanabhan, arXiv:0807.2356  (2008).


\bibitem{ccreview}
  R.~Bousso,
  Gen.\ Rel.\ Grav.\  {\bf 40}, 607-637 (2008).
  [arXiv:0708.4231 [hep-th]].


\bibitem{Gong:2006gs}
Y.-G. Gong and A. Wang, Phys. Rev. {\bf D75},  043520  (2007).

\bibitem{Komatsu:2008hk}
E. Komatsu {\it et~al.}, Astrophys. J. Suppl. {\bf 180},  330  (2009).

\bibitem{mycoin}
J. Lee, H.-C. Kim, and J.-W. Lee, Phys. Lett. B {\bf 661},  67  (2007).

\bibitem{myquantum}
  J.~-W.~Lee,
  Found.\ Phys.\  {\bf 41}, 744 (2011)  [arXiv:1005.2739 [hep-th]].  


\end{thebibliography}

\end{document}